\documentclass[showpacs,10pt,twocolumn,prb]{revtex4-1}

\usepackage{amsmath}
\usepackage{amssymb}
\usepackage{graphicx}
\usepackage{amssymb}
\usepackage{graphics}
\usepackage{epsfig}
\usepackage{CJK}
\usepackage{color}

\begin{document}

\begin{CJK*}{GBK}{}
\title{Critical behavior of quasi-two-dimensional semiconducting ferromagnet CrGeTe$_3$}
\author{Yu Liu,$^{1}$ and C. Petrovic$^{1}$}
\affiliation{$^{1}$Condensed Matter Physics and Materials Science Department, Brookhaven National Laboratory, Upton, New York 11973, USA}
\date{\today}

\begin{abstract}
The critical properties of the single-crystalline semiconducting ferromagnet CrGeTe$_3$ were investigated by bulk dc magnetization around the paramagnetic to ferromagnetic phase transition. Critical exponents $\beta = 0.200\pm0.003$ with critical temperature $T_c = 62.65\pm0.07$ K and $\gamma = 1.28\pm0.03$ with $T_c = 62.75\pm0.06$ K are obtained by the Kouvel-Fisher method whereas $\delta = 7.96\pm0.01$ is obtained by the critical isotherm analysis at $T_c = 62.7$ K. These critical exponents obey the Widom scaling relation $\delta = 1+\gamma/\beta$, indicating self-consistency of the obtained values. With these critical exponents the isotherm $M(H)$ curves below and above the critical temperatures collapse into two independent universal branches, obeying the single scaling equation $m = f_\pm(h)$, where $m$ and $h$ are renormalized magnetization and field, respectively. The determined exponents match well with those calculated from the results of renormalization group approach for a two-dimensional Ising system coupled with long-range interaction between spins decaying as $J(r)\approx r^{-(d+\sigma)}$ with $\sigma=1.52$.
\end{abstract}

\maketitle
\end{CJK*}

\section{INTRODUCTION}

Two-dimensional (2D) materials have recently stimulated significant attention not only for the emergence of novel properties but also for the potential applications.\cite{Wang, Geim, Lebegue, Gong, Ji} Particularly, layered intrinsically ferromagnetic (FM) semiconductors are of great interest since both ferromagnetism and semiconducting character are of interest for the next-generation spintronic devices.\cite{Sachs, Yamada,  Kabbour, McGuire, Casto, Zhang} CrXTe$_3$ (X = Si, Ge) crystals belong to this class; they have a band gap of 0.4 eV for CrSiTe$_3$ or 0.7 eV for CrGeTe$_3$, and simultaneously, exhibit ferromagnetic ordering below the Curie temperature ($T_c$) of 32 K for CrSiTe$_3$ or 61 K for CrGeTe$_3$, respectively.\cite{Casto, Zhang, Carteaux1, Carteaux2, Ouvrard, Siberchicot}

Considerable effort has been devoted in order to shed light on the nature of ferromagnetism in CrXTe$_3$, in particular the monolayer properties are of interest.\cite{Li, Chen, Sivadas, Liu, Lin} Previous neutron scattering showed that bulk CrSiTe$_3$ is a strongly anisotropic 2D Ising-like ferromagnet with a critical exponent $\beta = 0.17$ and a spin gap of $\sim$ 6 meV.\cite{Carteaux3} The critical behavior of CrSiTe$_3$ investigated by bulk magnetization measurements further confirms the critical exponent $\beta = 0.170\pm0.008$, comparable to $\beta = 0.125$ for a 2D Ising model.\cite{BJLiu} However, the recent neutron work on CrSiTe$_3$ observed $\beta = 0.151$ and a very small spin gap of $\sim$ 0.075 meV.\cite{Williams} Based on the spin wave analysis, the spins in CrSiTe$_3$ are Heisenberg-like.\cite{Williams} The spin wave theory suggests also that CrGeTe$_3$ is a nearly ideal 2D Heisenberg ferromagnet.\cite{Gong} The Monte Carlo simulations based on a Heisenberg model predict that the robust 2D ferromagnetism exists in nano-sheets of a single CrXTe$_3$ layer with $T_c$ $\sim$ 35.7 K for CrSiTe$_3$ or $\sim$ 57.2 K for CrGeTe$_3$.\cite{Li} By applying a moderate tensile strain, the 2D ferromagnetism can be largely enhanced with $T_c$ increasing to $\sim$ 91.7 K for CrSiTe$_3$ or $\sim$ 108.9 K for CrGeTe$_3$, respectively.\cite{Li} However, the Mermin-Wanger theorem states that long-range ferromagnetic order should not exist at non-zero temperature based on a 2D isotropic Heisenberg model,\cite{Mermin} with the exception of that the spins in the 2D system are constrained to only one direction, i.e., Ising-like spins.\cite{Zhuang}

When the second and third nearest-neighbor (NN) exchange interactions are considered, the monolayer CrSiTe$_3$ is expected to be an antiferromagnet with a zigzag spin texture whereas CrGeTe$_3$ is still a ferromagnet with $T_c$ of 106 K.\cite{Sivadas} This is in contrast with the result for ferromagnet where only the NN exchange interaction was considered.\cite{Sivadas}  An uniform in-plane tensile strain of $\sim$ 3$\%$ can tune the ground state of CrSiTe$_3$ from zigzag to ferromagnet with $T_c$ of 111 K.\cite{Sivadas}

In order to clarify the magnetic behavior in few-layer samples and the possible applications of this material, it is necessary to establish the nature of the magnetism in the bulk. In this paper, we investigated the critical behavior of CrGeTe$_3$ by various techniques, such as modified Arrott plot, Kouvel-Fisher plot, and critical isotherm analysis. Our analysis indicate that the obtained critical exponents $\beta = 0.200\pm0.003$ ($T_c = 62.65\pm0.07$ K), $\gamma = 1.28\pm0.03$ ($T_c = 62.75\pm0.06$ K), and $\delta = 7.96\pm0.01$ ($T_c = 62.7$ K) are in good agreement with those calculated from the results of renormalization group approach for 2D Ising model coupled with long-range interaction between spins decaying as $J(r)\approx r^{-(d+\sigma)}$ with $\sigma=1.52$.

\begin{figure}
\centerline{\includegraphics[scale=1.06]{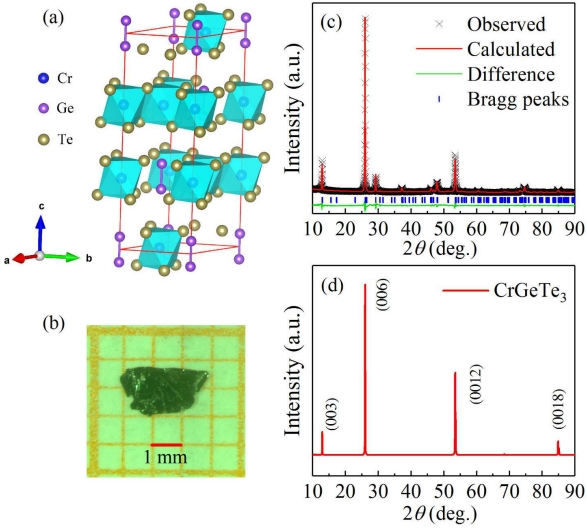}}
\caption{(Color online). (a) Crystal structure of CrGeTe$_3$. (b) Image of a representative single-crystalline sample. (c) Powder x-ray diffraction (XRD) and (d) single-crystal XRD pattern of CrGeTe$_3$. The vertical tick marks represent Bragg reflections of the $R\bar{3}h$ space group.}
\label{XRD}
\end{figure}

\section{EXPERIMENTAL DETAILS}

High quality CrGeTe$_3$ single crystals were grown by the self-flux technique starting from an intimate mixture of pure elements Cr (99.95 $\%$, Alfa Aesar) powder, Ge (99.999 $\%$, Alfa Aesar) pieces, and Te (99.9999 $\%$, Alfa Aesar) pieces with a molar ratio of 1 : 2 : 6. The starting materials were sealed in an evacuated quartz tube, which was heated to 1100 $^\circ$C over 20 h, held at 1100 $^\circ$C for 3 h, and then slowly cooled to 700 $^\circ$C at a rate of 1 $^\circ$C/h. X-ray diffraction (XRD) data were taken with Cu K$_{\alpha}$ ($\lambda=0.15418$ nm) radiation of Rigaku Miniflex powder diffractometer. The element analysis was performed using an energy-dispersive x-ray spectroscopy (EDX) in a JEOL LSM-6500 scanning electron microscope. The magnetization was measured in a Quantum Design Magnetic Property Measurement System (MPMS-XL5). Isotherms were collected at an interval of 0.5 K around $T_c$. The applied magnetic field ($H_a$) has been corrected for the internal field as $H = H_a - NM$, where $M$ is the measured magnetization and $N$ is the demagnetization factor. The corrected $H$ was used for the analysis of critical behavior.

\section{RESULTS AND DISCUSSIONS}

Figure 1(a) shows the crystal structure of bulk CrGeTe$_3$. Each unit cell comprises three CrGeTe$_3$ layers stacked in an ABC sequence along the $c$-axis. The Cr ions are located at the centers of slightly distorted octahedra of Te atoms. The Ge pairs form Ge$_2$Te$_6$ ethane-like groups. The as-grown single crystals are plate-like, typically 3 to 4 mm in size, as shown in Fig. 1(b). Figure 1(c) presents the powder x-ray diffraction (XRD) pattern of CrGeTe$_3$, in which the observed peaks are well fitted with the $R\bar{3}h$ space group. The determined lattice parameters are $a = 6.8263(3)$ {\AA} and $c = 20.5314(3)$ {\AA}, respectively. Furthermore, in the single crystal 2$\theta$ XRD scan [Fig. 1(d)], only $(00l)$ peaks are detected, indicating the crystal surface is normal to the $c$ axis with the plate-shaped surface parallel to the $ab$-plane.

\begin{figure}
\centerline{\includegraphics[scale=0.34]{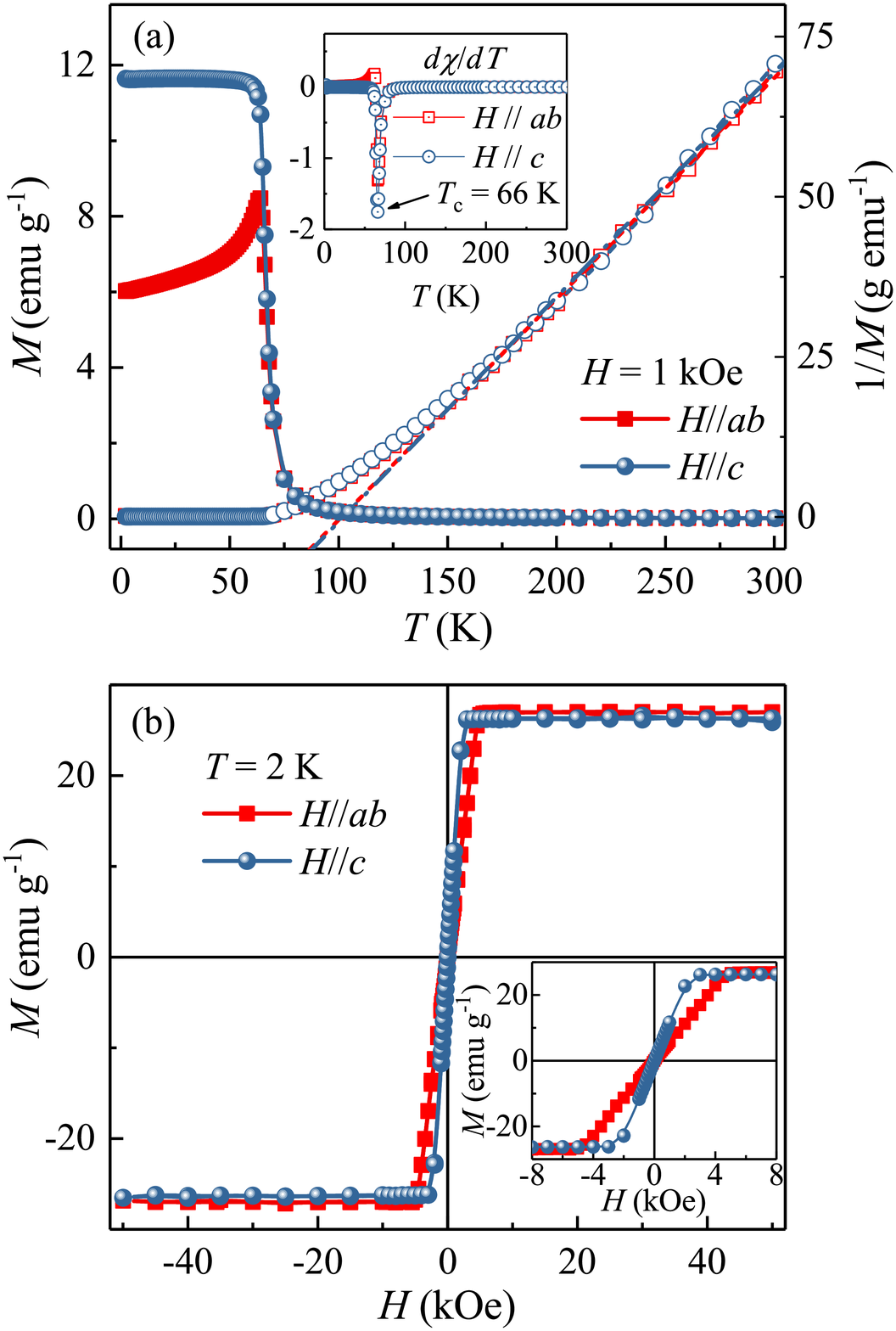}}
\caption{(Color online). (a) Temperature dependence of magnetization for CrGeTe$_3$ measured in the magnetic field $H$ = 1 kOe. Inset: the derivative magnetization $dM/dT$ versus $T$. (b) Field dependence of magnetization for CrGeTe$_3$ measured at $T$ = 2 K. Inset: the magnification of the low field region.}
\label{MTH}
\end{figure}

\begin{figure}
\centerline{\includegraphics[scale=0.366]{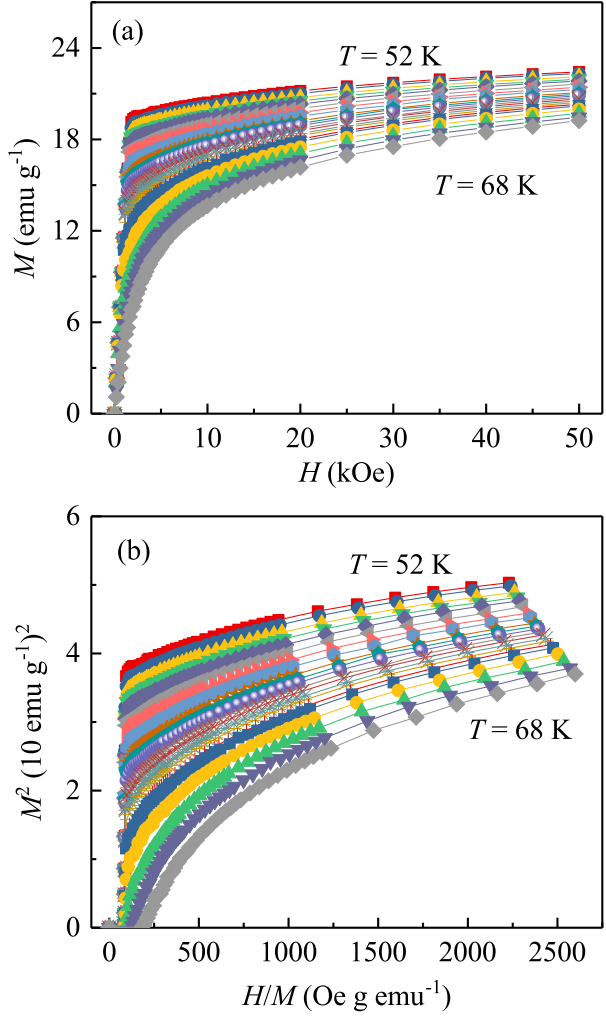}}
\caption{(Color online). (a) Typical initial isothermal magnetization curves from $T$ = 52 K to $T$ = 68 K for CrGeTe$_3$. (b) Arrott plots of $M^2$ versus $H/M$ around $T_c$ for CrGeTe$_3$.}
\label{Arrot}
\end{figure}

Figure 2(a) shows the temperature dependence of magnetization $M(T)$ measured in $H$ = 1 kOe applied in the $ab$-plane and parallel to $c$-axis, respectively. A clear paramagnetic (PM) to ferromagnetic (FM) transition is observed and the apparent anisotropy suggests that the crystallographic $c$-axis is the easy axis. As shown in the inset of Fig. 2(a), the critical temperature $T_c \approx 66$ K is roughly determined from the minimum of the $dM/dT$ curve. The temperature dependence of $1/M$ is also plotted in Fig. 2(a). A linear fit of the $1/M$ data in the temperature range of 150 to 300 K yields the Weiss temperature $\theta_{ab} \approx 108(1)$ K or $\theta_c \approx 113(2)$ K, which is nearly twice the value of $T_c$, indicating strong FM interaction in CrGeTe$_3$. The effective moment $\mu_{\textrm{eff}}$ = 3.43(2) $\mu_B$ obtained from $H//ab$ data is identical to $\mu_{\textrm{eff}}$ = 3.41(5) $\mu_B$ from $H//c$ data, which is close to the the theoretical value expected for Cr$^{3+}$ of 3.87 $\mu_B$. Figure 2(b) displays the isothermal magnetization measured at $T$ = 2 K. The saturation field $H_s \approx 3000 $ Oe for $H//c$ is smaller than $H_s \approx 5000 $ Oe for $H//ab$, confirming the easy axis is the $c$-axis. The saturation moment at $T$ = 2 K is $M_s \approx$ 2.45(1) $\mu_B$ for $H//ab$ and $M_s \approx$ 2.39(1) $\mu_B$ for $H//c$, respectively, close to the expected value of 3 $\mu_B$ for Cr$^{\textrm{+3}}$ with three unpaired spins. The inset of Fig. 2(b) shows the $M(H)$ in the low field region and the absence of coercive force ($H_c$) for CrGeTe$_3$.

\begin{figure}
\centerline{\includegraphics[scale=0.366]{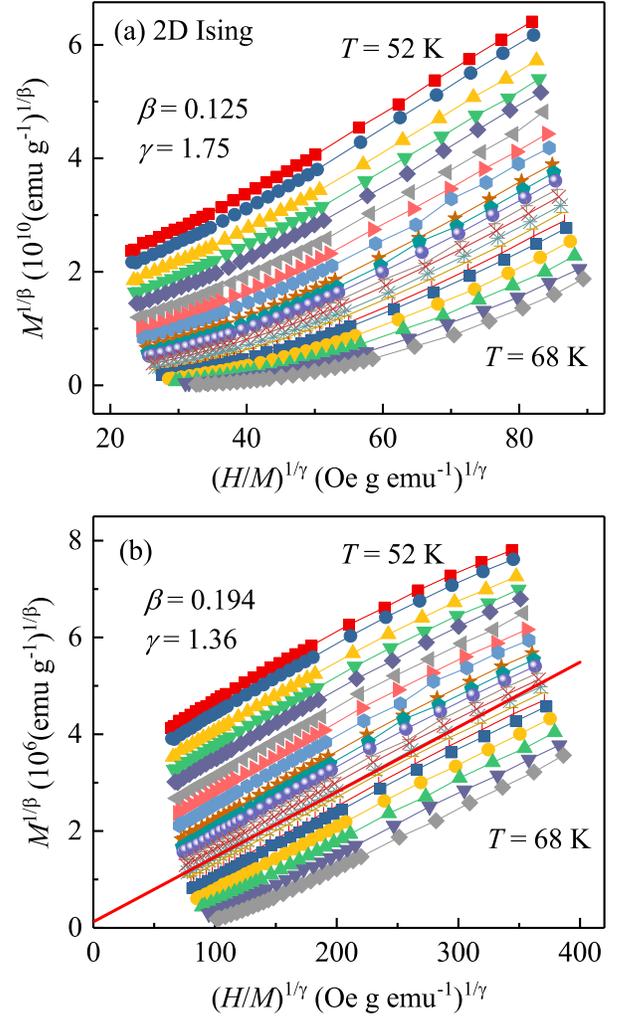}}
\caption{(Color online). (a) 2D-Ising model plot of isotherms for CrGeTe$_3$. (b) Modified Arrott plot of $M^{1/\beta}$ versus $(H/M)^{1/\gamma}$ with $\beta = 0.194$ and $\gamma = 1.36$ for CrGeTe$_3$. The straight line is the linear fit of isotherm at $T$ = 62.5 K.}
\label{Ising}
\end{figure}

\begin{figure}
\centerline{\includegraphics[scale=0.4]{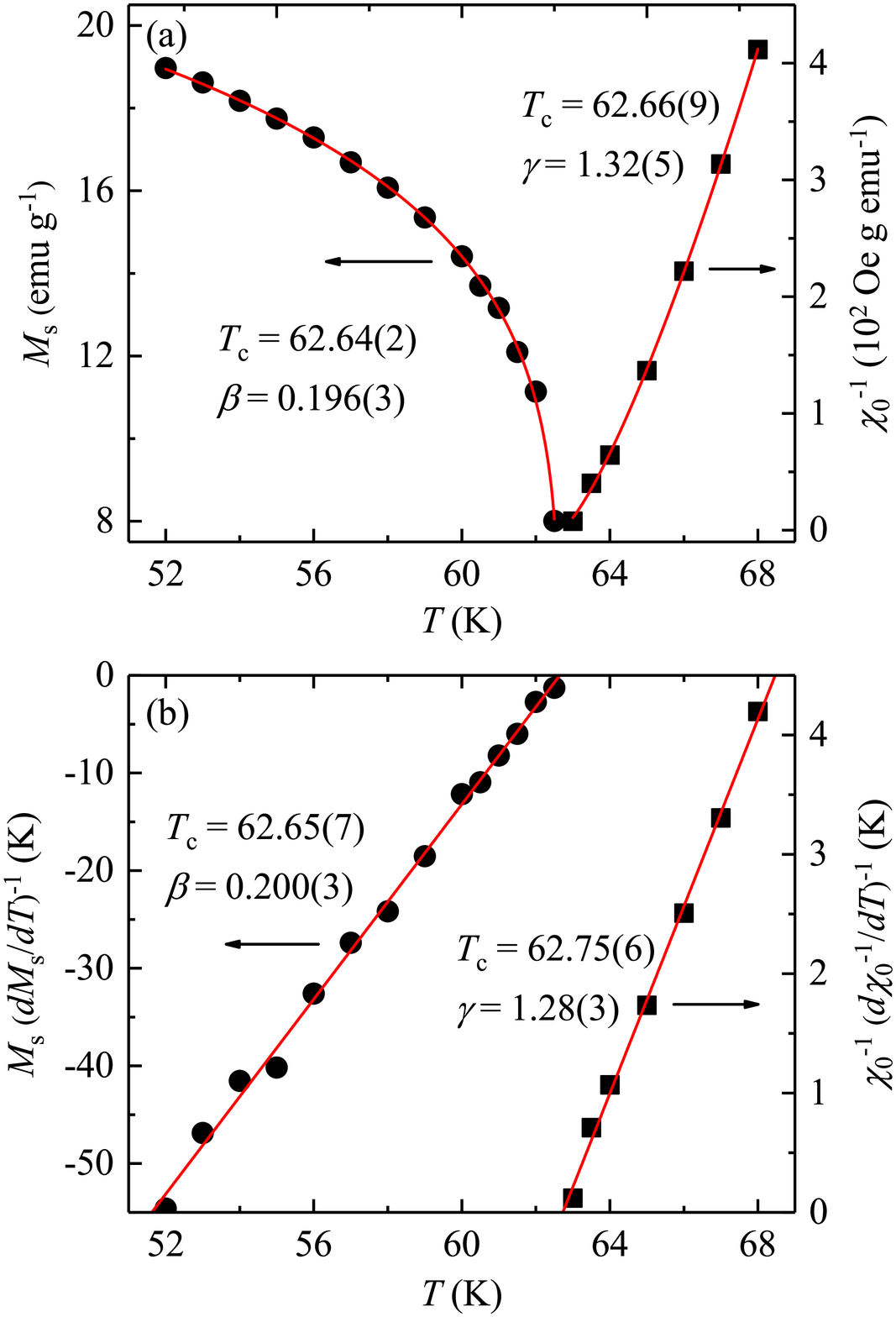}}
\caption{(Color online). (a) Temperature dependence of the spontaneous magnetization $M_s$ (left) and the inverse initial susceptibility $\chi_0^{-1}$ (right) with solid fitting curves for CrGeTe$_3$. (b) Kouvel-Fisher plots of $M_s(dM_s/dT)^{-1}$ (left) and $\chi_0^{-1}(d\chi_0^{-1}/dT)^{-1}$ (right) with solid fitting curves for CrGeTe$_3$.}
\label{KF}
\end{figure}

The critical behavior of a second-order transition can be characterized in detail by a series of interrelated critical exponents.\cite{Stanley} In the vicinity of a second-order phase transition, the divergence of correlation length $\xi = \xi_0 |(T-T_c)/T_c|^{-\nu}$ leads to universal scaling laws for the spontaneous magnetization $M_s$ and the inverse initial magnetic susceptibility $\chi_0^{-1}$. The spontaneous magnetization $M_s$ below $T_c$, the inverse initial susceptibility $\chi_0^{-1}$ above $T_c$, and the measured magnetization $M(H)$ at $T_c$ are characterized by a set of critical exponents $\beta$, $\gamma$, and $\delta$. The mathematical definitions of these exponents from magnetization are:
\begin{equation}
M_s (T) = M_0(-\varepsilon)^\beta, \varepsilon < 0, T < T_c
\end{equation}
\begin{equation}
\chi_0^{-1} (T) = (h_0/m_0)\varepsilon^\gamma, \varepsilon > 0, T > T_c
\end{equation}
\begin{equation}
M = DH^{1/\delta}, \varepsilon = 0, T = T_c
\end{equation}
where $\varepsilon = (T-T_c)/T_c$ is the reduced temperature, and $M_0$, $h_0/m_0$ and $D$ are the critical amplitudes.\cite{Fisher} The magnetic equation of state is a relationship among the variables $M(H,\varepsilon)$, $H$, and $T$. Using scaling hypothesis this can be expressed as:
\begin{equation}
M(H,\varepsilon) = \varepsilon^\beta f_\pm(H/\varepsilon^{\beta+\gamma})
\end{equation}
where $f_+$ for $T>T_c$ and $f_-$ for $T<T_c$, respectively, are the regular functions. In terms of renormalized magnetization $m\equiv\varepsilon^{-\beta}M(H,\varepsilon)$ and renormalized field $h\equiv\varepsilon^{-(\beta+\gamma)}H$, the Eq.(4) can be written as:
\begin{equation}
m = f_\pm(h)
\end{equation}
it implies that for true scaling relations and right choice of $\beta$, $\gamma$, and $\delta$ values, scaled $m$
and $h$ will fall on two universal curves: one above $T_c$ and another below $T_c$. This is an important criterion for the critical regime.

\begin{table*}
\caption{\label{tab}Comparison of critical exponents of CrGeTe$_3$ with different theoretical models.}
\begin{ruledtabular}
\begin{tabular}{llllll}
  Composition & Reference & Technique & $\beta$ & $\gamma$ & $\delta$ \\
  \hline
  CrGeTe$_3$ & This work & Modified Arrott plot & 0.196(3) & 1.32(5) & 7.73(15) \\
  & This work & Kouvel-Fisher plot & 0.200(3) & 1.28(3) & 7.40(5) \\
  & This work &Critical isotherm  &   &   & 7.96(1) \\
  2D Ising & 30 & Theory & 0.125 & 1.75 & 15 \\
  Mean field & 28 & Theory & 0.5 & 1.0 & 3.0 \\
  3D Heisenberg & 28 & Theory & 0.365 & 1.386 & 4.8 \\
  3D XY & 28 & Theory & 0.345 & 1.316 & 4.81 \\
  3D Ising & 28 & Theory & 0.325 & 1.24 & 4.82 \\
  Tricritical mean field & 29 & Theory & 0.25 & 1.0 & 5
\end{tabular}
\end{ruledtabular}
\end{table*}

In order to clarify the nature of PM-FM transition in CrGeTe$_3$, we measured the isothermal $M(H)$ in the temperature range from $T$ = 52 K to $T$ = 68 K, as shown in Fig. 3(a). Generally, conventional method to determine the critical exponents and critical temperature involves the use of Arrott plot.\cite{Arrott1} The Arrott plot assumes the critical exponents following the mean-field theory with $\beta$ = 0.5 and $\gamma$ = 1.0. According to this method, isotherms plotted in the form of $M^2$ versus $M/H$ constitute a set of parallel straight lines, and the isotherm at the critical temperature $T_c$ should pass through the origin. At the same time, it directly gives $\chi_0^{-1}(T)$ and $M_s(T)$ as the intercepts on $H/M$ axis and positive $M^2$ axis, respectively. Figure 3(b) shows the Arrott plot of CrGeTe$_3$. However, all the curves in this plot show nonlinear behavior having downward curvature even in high fields. This suggests that the mean-field model is not valid for CrGeTe$_3$. According to the Banerjee$^\prime$s criterion,\cite{Banerjee} one can estimate the order of the magnetic transition through the slope of the straight line: negative slope corresponds to the first-order transition while positive corresponds to the second-order. Therefore, the concave downward curvature clearly indicates the PM-FM transition in CrGeTe$_3$ is a second-order one.

Considering the strong two-dimensional (2D) characteristics in CrGeTe$_3$, we further analyze the isothermal data with 2D-Ising model ($\beta$ = 0.125, $\gamma$ = 1.75).\cite{LeGuillou} As shown in Fig. 4(a), a set of quasi-parallel straight lines are obtained. However, it still can not find a single straight line that passes through origin, indicating that CrGeTe$_3$ can not be rigorously described by the 2D-Ising model. Therefore, a modified Arrott plot by a self-consistent method are further applied to determine $T_c$ as well as the critical exponents $\beta$ and $\gamma$.\cite{Arrott2} The modified Arrott plot is given by the Arrot-Noaks equation of state:
\begin{equation}
(H/M)^{1/\gamma} = a\varepsilon+bM^{1/\beta}
\end{equation}
where $\varepsilon = (T-T_c)/T_c$ is the reduced temperature, $a$ and $b$ are constants. To find out the proper values of $\beta$ and $\gamma$, a rigorous iterative method has been used.\cite{Pramanik} The starting values of $M_s(T)$ and $\chi_0^{-1}(T)$ were determined from the 2D-Ising model plot by the linear extrapolation from the high field region to the intercepts with the axis $M^{1/\beta}$ and $(H/M)^{1/\gamma}$, respectively. A new set of $\beta$ and $\gamma$ can be obtained by fitting data following the Eqs (1) and (2). Then the obtained new values of $\beta$ and $\gamma$ are used to reconstruct a new modified Arrott plot. This procedure was repeated until the values of $\beta$ and $\gamma$ are stable. By this method, the obtained critical exponents are hardly dependent on the initial parameters, which confirms these critical exponents are reliable and intrinsic. The final modified Arrot plots generated with the values $\beta = 0.194$ and $\gamma = 1.36$ are depicted in Fig. 4(b).

Figure 5(a) presents the final $M_s(T)$ and $\chi_0^{-1}(T)$ with the solid fitting curves. The critical exponents $\beta = 0.196(3)$ with $T_c = 62.64(2)$ K and $\gamma = 1.32(5)$ with $T_c = 62.66(9)$ K are obtained, which are very close to the values obtained from the modified Arrot plot in Fig. 4(b). Alternatively, the critical exponents can be determined by the Kouvel-Fisher (KF) method:\cite{Kouvel}
\begin{equation}
\frac{M_s(T)}{dM_s(T)/dT} = \frac{T-T_c}{\beta}
\end{equation}
\begin{equation}
\frac{\chi_0^{-1}(T)}{d\chi_0^{-1}(T)/dT} = \frac{T-T_c}{\gamma}
\end{equation}
According to this method, $M_s(T)dM_s(T)/dT$ and $\chi_0^{-1}(T)d\chi_0^{-1}(T)/dT$ are as linear functions of temperature with slopes of $1/\beta$ and $1/\gamma$, respectively. As shown in Fig. 5(b), the linear fits give $\beta = 0.200(3)$ with $T_c = 62.65(7)$ K and $\gamma = 1.28(3)$ with $T_c = 62.75(6)$ K, respectively.

\begin{figure}
\centerline{\includegraphics[scale=0.4]{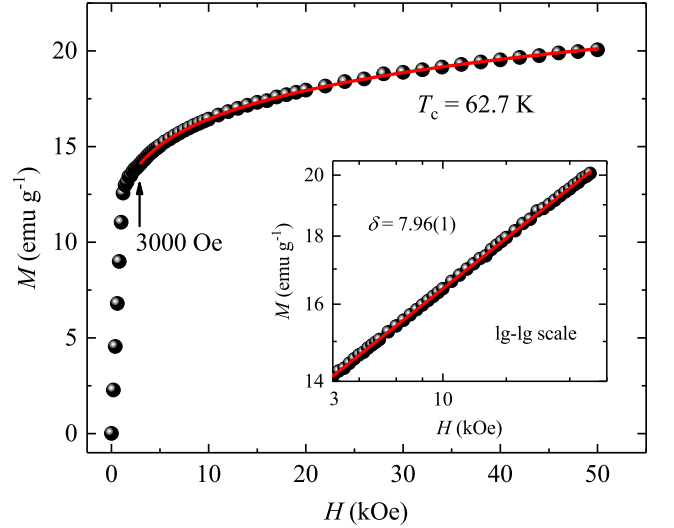}}
\caption{(Color online). Isotherm $M$ versus $H$ plot collected at $T_c$ = 62.7 K for CrGeTe$_3$. Inset: the same plot in log-log scale with a solid fitting curve.}
\label{MH}
\end{figure}

The isothermal magnetization $M(H)$ at the critical temperature $T_c$ = 62.7 K is shown in Fig. 6, with the inset plotted on a lg-lg scale. According to Eq. (3), the third critical exponent $\delta = 7.96(1)$ can be deduced. Furthermore, the exponent $\delta$ can also been calculated from Widom scaling relation according to which critical exponents $\beta$, $\gamma$, and $\delta$ are related in following way:
\begin{equation}
\delta = 1+\frac{\gamma}{\beta}
\end{equation}
Using the $\beta$ and $\gamma$ values determined from Modified Arrott plot and Kouvel-Fisher plot, we obtain $\delta$ = 7.73(15) and $\delta$ = 7.40(5), respectively, which are very close to the value obtained from critical isotherm analysis. Therefore, the critical exponents and $T_c$ obtained in present study are self-consistent and an accurate estimate within experimental precision.

All these critical exponents derived from various methods are given in Table~\ref{tab} along with the theoretically predicted values for different models. The reliability of the obtained critical exponents and $T_c$ can also be verified by a scaling analysis. Following Eq. (5), scaled $m$ versus scaled $h$ has been plotted in Fig. 7(a), along with the same plot on lg-lg scale in the inset of Fig. 7(a). It is rather significant that all the data collapse into two separate branches: one below $T_c$ and another above $T_c$. The reliability of the exponents and $T_c$ has been further ensured with more rigorous method by plotting $m^2$ versus $h/m$, as shown in Fig. 7(b), where all data also fall on two independent branches. This clearly indicates that the interactions get properly renormalized in critical regime following scaling equation of state. In addition, the scaling equation of state takes another form:
\begin{equation}
\frac{H}{M^\delta} = k(\frac{\varepsilon}{H^{1/\beta}})
\end{equation}
where $k(x)$ is the scaling function. Based on Eq. (10), all experimental curves will collapse onto a single curve. The inset of Fig. 7(b) shows the $MH^{-1/\delta}$ versus $\varepsilon H^{-1/(\beta\delta)}$ for CrGeTe$_3$, where the experimental data collapse onto a single curve, and $T_c$ locates at the zero point of the horizontal axis. The well-rescaled curves further confirm the reliability of the obtained critical exponents. As we can see, the experimentally determined critical exponents $\beta$, $\gamma$, and $\delta$ show some deviation from the theoretical values of 2D-Ising model, which might be associated with non-negligible interlayer coupling and spin-lattice coupling in this material.

\begin{figure}
\centerline{\includegraphics[scale=0.38]{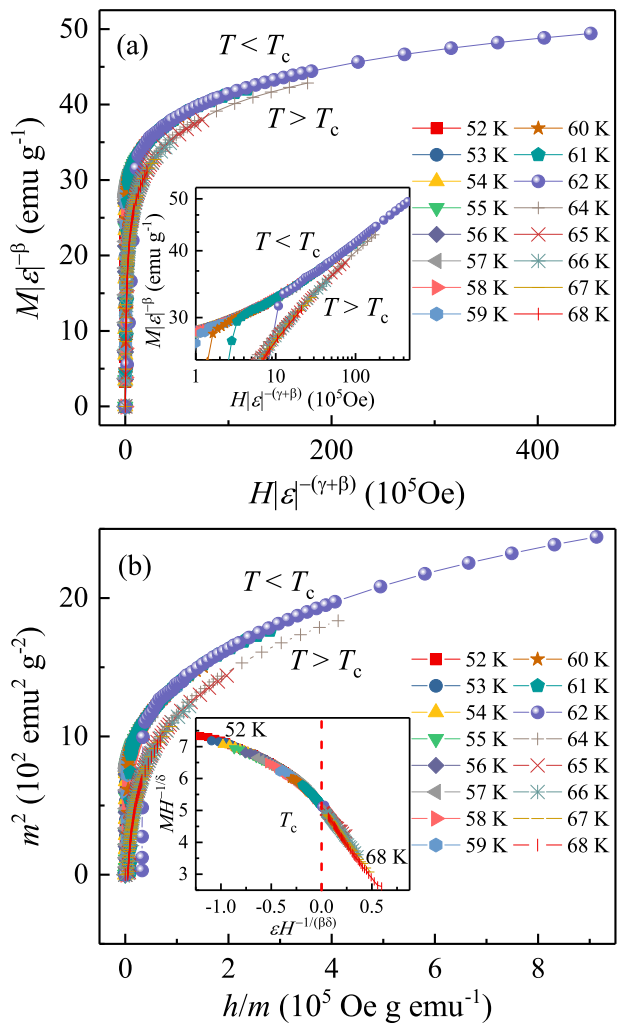}}
\caption{(Color online). (a) Scaling plots of renormalized magnetization $m$ versus renormalized field $h$ below and above $T_c$ for CrGeTe$_3$. Inset: the same plots in log-log scale. (b) the renormalized magnetization and field replotted in the form of $m^2$ versus $h/m$ for CrGeTe$_3$. Inset: the rescaling of the $M(H)$ curves by $MH^{-1/\delta}$ versus $\varepsilon H^{-1/(\beta\delta)}$.}
\label{magnetism}
\end{figure}

Finally, we would like to discuss the nature as well as the range of interaction in CrGeTe$_3$. For a homogeneous magnet, the universality class of the magnetic phase transition depending on the exchange distance $J(r)$. Fisher \emph{et al.} theoretically treated this kind of magnetic ordering as an attractive interaction of spins, where a renormalization group theory analysis suggests the interaction decays with distance $r$ as:
\begin{equation}
J(r) \approx r^{-(d+\sigma)}
\end{equation}
where $d$ is the spatial dimensionality and $\sigma$ is a positive constant.\cite{Fisher1972} According to this model, the range of the spin interaction is long or short depending on the $\sigma < 2$ or $\sigma > 2$, and it predicts the susceptibility exponent $\gamma$ which has been calculated from renormalization group approach, as following:
\begin{multline}
\gamma = 1+\frac{4}{d}(\frac{n+2}{n+8})\Delta\sigma+\frac{8(n+2)(n-4)}{d^2(n+8)^2}\\\times[1+\frac{2G(\frac{d}{2})(7n+20)}{(n-4)(n+8)}]\Delta\sigma^2
\end{multline}
where $\Delta\sigma = (\sigma-\frac{d}{2})$ and $G(\frac{d}{2})=3-\frac{1}{4}(\frac{d}{2})^2$.
To find out the range of interaction ($\sigma$) as well as the dimensionality of both lattice ($d$) and spin ($n$) in this system we have followed the procedure similar to Ref. 35 where the parameter $\sigma$ in above expression is adjusted for a particular values of \{$d:n$\} so that it yields a value for $\gamma$ close to that experimentally observed $\gamma = 1.28$. The so obtained $\sigma$ is then used to calculate the remaining exponents from the following expressions: $\nu = \gamma/\sigma$, $\alpha = 2-\nu d$, $\beta = (2-\alpha-\gamma)/2$, and $\delta = 1+\gamma/\beta$. This exercise is repeated for different set of \{$d:n$\}. We found that \{$d:n$\} = \{2:1\} and $\sigma = 1.52$ give the exponents ($\beta = 0.256$, $\gamma = 1.617$, and $\delta = 7.32$) which are close to our experimentally observed values (Table I). This calculation suggests the spin interaction in CrGeTe$_3$ is of 2D Ising (\{$d:n$\} = \{2:1\}) type coupled with long-range ($\sigma = 1.52$) interaction.

\section{CONCLUSIONS}

In summary, we have made a comprehensive study on the critical phenomenon at the PM-FM phase transition in the quasi-2D semiconducting ferromagnet CrGeTe$_3$. This transition is identified to be second order in nature. The critical exponents $\beta$, $\gamma$, and $\delta$ estimated from various techniques match reasonably well and follow the scaling equation, confirming that the obtained exponents are unambiguous and intrinsic to the material. The determined exponents match well with those calculated from the results of renormalization group approach for a 2D Ising (\{$d:n$\} = \{2:1\}) system coupled with long-range interaction between spins decaying as $J(r)\approx r^{-(d+\sigma)}$ with $\sigma=1.52$.
\textit{Note added}. Recently, we became aware that G. T. Lin\cite{LinGT} also synthesized CrGeTe$_{3}$. Their conclusions regarding tricritical point are not in conflict with our work.

\section*{Acknowledgements}
We thank John Warren for help with SEM measurements. This work was supported by the U.S. DOE-BES, Division of Materials Science and Engineering, under Contract No. DE-SC0012704 (BNL)


\begin{references}

\bibitem{Wang} Q. H. Wang, K. Kalantar-Zadeh, A. Kis, J. N. Coleman, and M. S. Strano, Nat. Nanotechnol., \textbf{7}, 699 (2012).
\bibitem{Geim} A. K. Geim, and I. V. Grigorieva, Nature, \textbf{499}, 419 (2013).
\bibitem{Lebegue} S. Leb\`{e}gue, T. Bj\"{o}rkman, M. Klintenberg, R. M. Nieminen, and O. Eriksson, Phys. Rev. X, \textbf{3}, 031002 (2013).
\bibitem{Gong} C. Gong, L. Li, Z. L. Li, H. W. Ji, A. Stern, Y. Xia, T. Cao, W. Bao, C. Z. Wang, Y. Wang, Z. Q. Qiu, R. J. Cava, S. G. Louie, J. Xia, and X. Zhang, Nature, doi:10.1038/nature22060 (2017).
\bibitem{Ji} H. W. Ji, R. A. Stokes, L. D. Aegria, E. C. Blomberg, M. A. Tanatar, A. Reijnders, L. M. Schoop, T. Liang, R. Prozorov, K. S. Burch, N. P. Ong, J. R. Petta, and R. J. Cava, J. Appl. Phys., \textbf{114}, 114907 (2013).
\bibitem{Sachs} B. Sachs, T. O. Wehling, K. S. Novoselov, A. I. Lichtenstein, and M. I. Katsnelson, Rhys. Rev. B, \textbf{88}, 201402 (2013).
\bibitem{Yamada} I. Yamada, J. Phys. Soc. Jpn., \textbf{33}, 979 (1972).
\bibitem{Kabbour} H. Kabbour, R. David, A. Pautrat, H. J. Koo, M. H. Whangbo, G. Andr\'{e}, and O. Mentr\'{e}, Angew. Chem., Int. Ed., \textbf{51}, 11745 (2012).
\bibitem{McGuire} M. A. McGuire, H. Dixit, V. R. Cooper, and B. C. Sales, Chem. Mater., \textbf{27}, 612 (2015).
\bibitem{Casto} L. D. Casto, A. J. Clune, M. O. Yokosuk, J. L. Musfeldt, T. J. Williams, H. L. Zhuang, M. W. Lin, K. Xiao, R. G. Hennig, B. C. Sales, J. Q. Yan, and D. Mandrus, APL Mater., \textbf{3}, 041515 (2015).
\bibitem{Zhang} X. Zhang, Y. L. Zhao, Q. Song, S. Jia, J. Shi, and W. Han, Jpn. J. Appl. Phys., \textbf{55}, 033001 (2016).
\bibitem{Carteaux1} V. Carteaux, G. Ouvrard, J. C. Grenier, and Y. Laligant, J. Magn. Magn. Mater., \textbf{94}, 127 (1991).
\bibitem{Carteaux2} V. Carteaux, D. Brunet, G. Ouvrard, and G. Andr\'{e}, J. Phys.: Condens. Matter, \textbf{7}, 69 (1995).
\bibitem{Ouvrard} G. Ouvrard, E. Sandre, and R. Brec, J. Solid State Chem., \textbf{73}, 27 (1988).
\bibitem{Siberchicot} B. Siberchicot, S. Jobic, V. Carteaux, P. Gressier, and G. Ouvrard, J. Phys. Chem., \textbf{100}, 5863 (1996).
\bibitem{Li} X. X. Li, and J. L. Yang, J. Mater. Chem. C, \textbf{2}, 7071 (2014).
\bibitem{Chen} X. F. Chen, J. S. Qi, and D. N. Shi, Phys. Lett. A, \textbf{379}, 60 (2015).
\bibitem{Sivadas} N. Sivadas, M. W. Daniels, R. H. Swendsen, S. Okamoto, and D. Xiao, Phys. Rev. B, \textbf{91}, 235425 (2015).
\bibitem{Liu} J. P. Liu, S. Y. Park, K. F. Garrity, and D. Vanderbilt, Phys. Rev. Lett., \textbf{117}, 257201 (2016).
\bibitem{Lin} M. W. Lin, H. L. Zhuang, J. Q. Yan, T. Z. Ward, A. A. Puretzky, C. M. Rouleau, Z. Gai, L. B. Liang, V. Meunier, B. G. Sumpter, P. Ganesh, P. R. C. Kent, D. B. Geohegan, D. G. Mandrus, and K. Xiao, J. Mater. Chem. C, \textbf{4}, 315 (2016).
\bibitem{Carteaux3} V. Carteaux, F. Moussa, and M. Spiesser, Europhys. Lett., \textbf{29}, 251 (1995).
\bibitem{BJLiu} B. J. Liu, Y. M. Zou, L. Zhang, S. M. Zhou, Z. Wang, W. K. Wang, Z. Qu, and Y. H. Zhang, Sci. Rep., \textbf{6}, 33873 (2016).
\bibitem{Williams} T. J. Williams, A. A. Aczel, M. D. Lumsden, S. E. Nagler, M. B. Stone, J. Q. Yan, and D. Mandrus, Phys. Rev. B, \textbf{92}, 144404 (2015).
\bibitem{Mermin} N. D. Mermin, and H. Wagner, Phys. Rev. Lett., \textbf{17}, 1133 (1966).
\bibitem{Zhuang} H. L. Zhuang, Y. Xie, P. R. C. Kent, and P. Ganesh, Phys. Rev. B, \textbf{92}, 035407 (2015).
\bibitem{Stanley} H. E. Stanley, \emph{Introduction to Phase Transitions and Critical Phenomena} (Oxford U. P., London and New York, 1971).
\bibitem{Fisher} M. E. Fisher, Rep. Prog. Phys., \textbf{30}, 615 (1967).
\bibitem{Arrott1} A. Arrott, Phys. Rev. B, \textbf{108}, 1394 (1957).
\bibitem{Banerjee} S. K. Banerjee, Phys. Lett., \textbf{12}, 16 (1964).
\bibitem{LeGuillou} J. C. LeGuillou, and J. Zinn-Justin, Phys. Rev. B, \textbf{21}, 3976 (1980).
\bibitem{Arrott2} A. Arrott, and J. Noakes, Phys. Rev. Lett., \textbf{19}, 786 (1967).
\bibitem{Pramanik} A. K. Pramanik, and A. Banerjee, Phys. Rev. B, \textbf{79}, 214426 (2009).
\bibitem{Kouvel} J. S. Kouvel, and M. E. Fisher, Phys. Rev., \textbf{136}, A1626 (1964).
\bibitem{Fisher1972} M. E. Fisher, S. K. Ma, and B. G. Nickel, Phys. Rev. Lett., \textbf{29}, 917 (1972).
\bibitem{Fischer} S. F. Fischer, S. N. Kaul, and H. Kronmuller, Phys. Rev. B, \textbf{65}, 064443 (2002).
\bibitem{LinGT} G. T. Lin, H. L. Zhuang, X. Luo, B.J. Liu, F. C. Chen, J. Yan, Y. Sun, J. Zhou, W. J. Lu, P. Tong, Z. G. Sheng, Z. Qu, W. H. Song, X. B. Zhu, and Y. P. Sun, arXiv:1706.03239.

\end{references}
\end{document}